\documentclass[conference]{IEEEtran}
\usepackage{graphicx}
\usepackage{amsmath}
\usepackage{amsfonts}%
\usepackage{amssymb}
\usepackage{epsfig}
\usepackage{algorithm}
\usepackage{algorithmic}
\usepackage{color}
\usepackage{array}
\usepackage{subcaption}
\usepackage{adjustbox}

\usepackage{makecell}

\ifCLASSINFOpdf
\else
\fi
\usepackage{amsmath}
\begin{document}
%
\title{Centralized and Distributed Machine Learning-Based QoT Estimation for Sliceable Optical Networks}
\author{\IEEEauthorblockN{Tania Panayiotou$^1$, Giannis Savva$^1$, Ioannis Tomkos$^2$, and Georgios Ellinas$^1$}
\IEEEauthorblockA{$^1$ KIOS Research and Innovation Center of Excellence,\\
Department of Electrical and Computer Engineering, 
University of Cyprus\\
Email: \{panayiotou.tania, gsavva07, gellinas\}@ucy.ac.cy}
$^2$ Athens Information Technology, Marousi, 15125 Athens, Greece,\\
Email: itom@ait.gr \\
}

%


\maketitle

\begin{abstract}
Dynamic network slicing has emerged as a promising and fundamental framework for meeting 5G's diverse use cases. As machine learning (ML) is expected to play a pivotal role in the efficient control and management of these networks, in this work we examine the ML-based Quality-of-Transmission (QoT) estimation problem under the dynamic network slicing context, where each slice has to meet a different QoT requirement. We examine ML-based QoT frameworks with the aim of finding QoT model/s that are fine-tuned according to the diverse QoT requirements. Centralized and distributed frameworks are examined and compared according to their accuracy and training time. We show that the distributed QoT models outperform the centralized QoT model, especially as the number of diverse QoT requirements increases.              
\end{abstract}


%
\IEEEpeerreviewmaketitle

\section{Introduction}
The fifth generation (5G) of mobile and wireless networks is expected to support a wide range of services (e.g., e-health, connected cars) and applications (e.g., video streaming), with very diverse characteristics and requirements (e.g., different bit rates, latency, and availability requirements). Dynamic network slicing has emerged as a promising and fundamental framework for meeting 5G's diverse use cases, including future-proof scalability and flexibility~\cite{1}. In general, network slicing is a form of virtual network architecture using the same principles as software defined networking (SDN) and network functions virtualization (NFV). SDN and NFV (currently deployed only at the edge of transport networks) will allow traditional network architectures to be dynamically (on demand) partitioned into virtual elements that can be also linked (through software). This will allow multiple virtual networks (slices) to be created on top of a common shared physical infrastructure with the slices being customized to meet the specific needs of each tenant~\cite{1,2}. Therefore, network slicing  has in the last few years been adopted as an emerging technology and business model by many telecom operators, such as, Ericsson, Nokia, and AT$\&$T~\cite{2,3}.

Although the implementation of network slicing is conceived to be an end-to-end feature, spanning over different technology domains (e.g., access, metro, and core networks), the attention, until recently, has been mainly focused on the radio access network (RAN)~\cite{1,2,4,5}. Recent work, however, focuses on extending the scope of dynamic slicing beyond the RAN to include both mobile and transport network segments (metro/core optical segments)~\cite{6}-\cite{8}. Specifically, existing work mainly focuses on demonstrating SDN/NFV and network orchestration implementations along with applying analytics on traffic demand patterns for addressing the vision of end-to-end network slicing. Undoubtedly, SDN/NFV, network orchestration, and analytics, constitute the key ingredients for achieving this objective~\cite{1}. However, little has been done so far regarding the efficient control and management of these networks, that will enable not only the creation of these slices/services but also the smart deployment of these slices to the right place at the right time with optimized resources.

It is expected that machine learning (ML) will play a pivotal role in the efficient control and management of these networks, enabling automation through the softwarization of the network planning functions that are becoming increasingly complex in an ever changing, heterogeneous, and uncertain environment. Several ML techniques have already been applied~\cite{12,13} that by and large aim to perform traffic prediction/estimation~\cite{17}-\cite{21}, fault detection/localization~\cite{26}-\cite{28b}, attack detection/identification~\cite{28c}, and QoT estimation~\cite{22}-\cite{25b}. Existing works on the QoT estimation problem, that this work also focuses on, aim at finding a QoT model that is fine tuned according to a single QoT requirement that reflects the highest QoT requirement amongst all services. Even though the single QoT requirement assumption perfectly fits the existing network planning scenario, where a single-slice has to best fit all the diverse QoT requirements, as metro/core optical networks are already transforming to a multi-slicing scenario~\cite{6}-\cite{8}, the model inference procedure has to also be transformed accordingly. The motivation behind the exploration of different QoT requirements in optical networks segments is based on the fact that next generation optical networks are envisioned/expected to be able to support services with various optical service level agreements (OSLAs), with the BER (QoT requirement) being amongst the OSLA specifications~\cite{15b}. 
 
On this basis, we examine centralized and distributed ML-based QoT estimation frameworks for sliceable optical networks under a multi-slicing scenario. The objective is to find QoT estimation model/s that are fine-tuned to the diverse QoT requirements of each slice. The centralized QoT estimation problem is formulated as a multiclass classifier that is trained according to global network information. The distributed QoT estimation problem is formulated as a set of binary classifiers with each classifier being trained independently according to information that is relevant only to a single type of slice; that is, a slice that consists of connections with the same QoT requirements. Both the centralized and distributed frameworks constitute novel ML-based QoT estimation approaches and their advantages as well as limitations are for the first time examined. We show that the distributed QoT estimation models outperform the centralized QoT model in both accuracy and training time, especially as the number of diverse slices increases.   

\section{Related Work}
\begin{figure}[h]
\begin{center}
\includegraphics[scale=0.25]{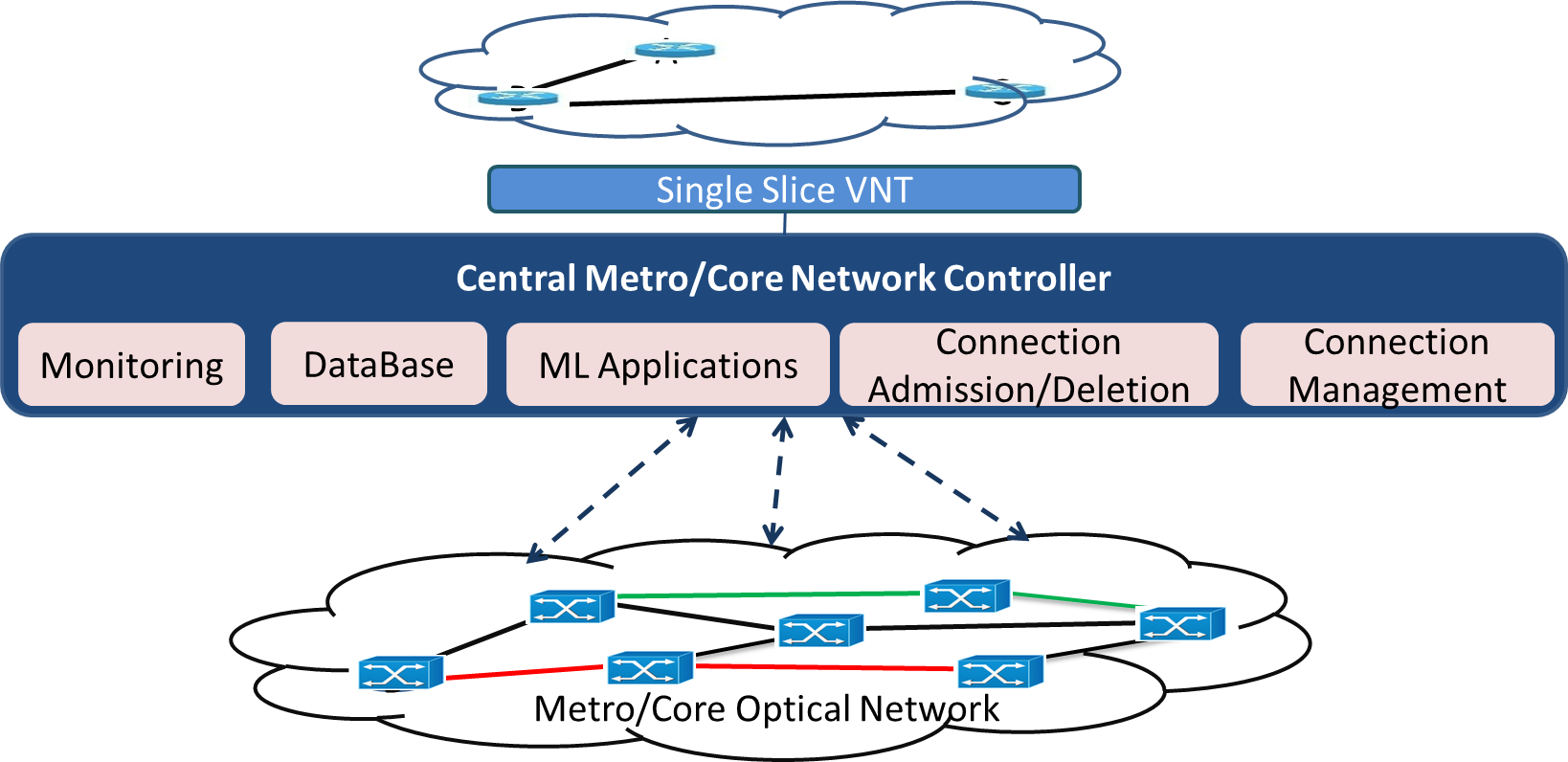}
\caption{Single-slice Control and Management Framework.}
\label{f1}
\end{center}
\end{figure}

Several ML applications have been already developed and explored for optical network planning purposes~\cite{12,13}. In general, the state-of-the-art assumes that the optical network is centrally controlled by an SDN-based optical network controller~\cite{17,15}, equipped with storage, processing, and monitoring capabilities (Fig.~\ref{f1}). The main responsibility of the SDN-based controller is to efficiently manage the network resources in such a way that the diverse QoS requirements of the different use cases are met, as closely as possible, by the virtual network topology (VNT). The VNT can be viewed as a single slice (virtual network) that has to best fit all the diverse use cases.   

To achieve this, ML applications run centrally, according to the monitored information (e.g., BER) that is stored in a central knowledge database. An ML application runs on top of the database, from which it extracts and analyzes such data, aiming to infer near-optimal/accurate ML models. Each ML application infers (usually) a single model (e.g., QoT model) that is subsequently used for centrally managing the network resources. Even though these ML models can be adapted to network changes (e.g., by retraining the models according to the most recent information), the single slice (single VNT) assumption leads to ML-based models that roughly meet the diverse QoS requirements of the different use cases (services). In fact, these models, in order to meet the QoS requirements of the diverse use cases, have to be trained according to the use-case/service having the highest QoS requirements. 

In particular, for the ML-based QoT estimation problem that this work focuses on, the QoT estimation model must be trained according to the use-case/service requiring the lowest BER, consequently leading to the underutilization of the network resources. Specifically, the state-of-the-art QoT estimation problem is usually formulated as a binary classifier based on a single BER threshold capable of ensuring that the QoT of each diverse connection is sufficient~\cite{22}-\cite{24}. Hence, a state-of-the-art ML-based QoT estimation model is, after the inference procedure, capable of classifying the unestablished lightpaths into one of two classes; the infeasible QoT class or the feasible QoT class~\cite{22}-\cite{24}. Since the feasible QoT class may be based/mapped to a BER threshold that is significantly lower than the true BER requirement of some services, the QoT model may lead to inefficient utilization of the network resources. To alleviate this problem, we examine alternative centralized and distributed ML-based frameworks that specifically take into account the BER requirements of the diverse services/slices. 

It is important to note that until now, and under the existing single-slicing approach, the forward error correction (FEC) schemes currently applied to the optical transport segment are designed to tolerate a single BER requirement (the lowest acceptable amongst all the services). However, for the adoption of a truly efficient end-to-end network slicing implementation, the FEC schemes will also need to be transformed accordingly. Such practical feasibility issues are out of the scope of this work and remain open for future research efforts. 

\section{Problem Statement}
We assume a sliceable optical metro/core network that is dynamically partitioned into a number of temporal virtual networks (slices) that meet the specific QoT requirements of each service/slice. We assume that in a metro/core network, a slice, unlike fronthaul networks, is not directly associated with a particular application (e.g., video streaming), but it is rather defined according to a set of QoS requirements of the same type. In these networks, slice (re)configuration is performed dynamically as connection requests arrive into the optical network. A connection $i$ is defined according to the set $C^i=\{s,d, B_k\}$, where $s$ is the source node, $d$ is the destination node, and $B_k$ is the BER requirement of connection $i$ for the slice type $k$, where $k=1,..,K$. Note that the $B_k$ requirement of each possible slice type can be known a priori (i.e., defined according to the OSLAs that are set between the network operators and the end-users). Hence, a connection request can be provisioned according to the set $C^i$. In an elastic optical network (EON) this can be achieved by solving the QoT-aware routing and spectrum allocation problem. This requires the existence of predefined QoT estimation model/s for ensuring that each computed lightpath will meet its QoT requirement, before it is actually established into the network.  

On this basis, the objective is to find accurate QoT models that are fine tuned according to the BER requirements of each slice type $k$. These models can then be used during the slice provisioning phase for ensuring that the QoT requirements of each slice type will be met before these slices are actually (re)configured. For finding these models, the QoT estimation problem is formulated according to both the ML-based centralized-computed framework and the ML-based distributed-computed framework analytically discussed in Section~\ref{frame} (Fig.~\ref{f2}). Briefly, in the centralized approach, a single model is inferred in a central controller according to a multiclass classifier serving all the slice types, while in the distributed approach, a set of binary classifiers are inferred in distributed controllers, with each classifier serving a single slice type. Both frameworks are examined and compared according to their models' accuracy and training time.  
  
\section{QoT Estimation Frameworks in Sliceable Optical Networks}~\label{frame} 
\begin{figure}[h]
\begin{center}
\includegraphics[scale=0.23]{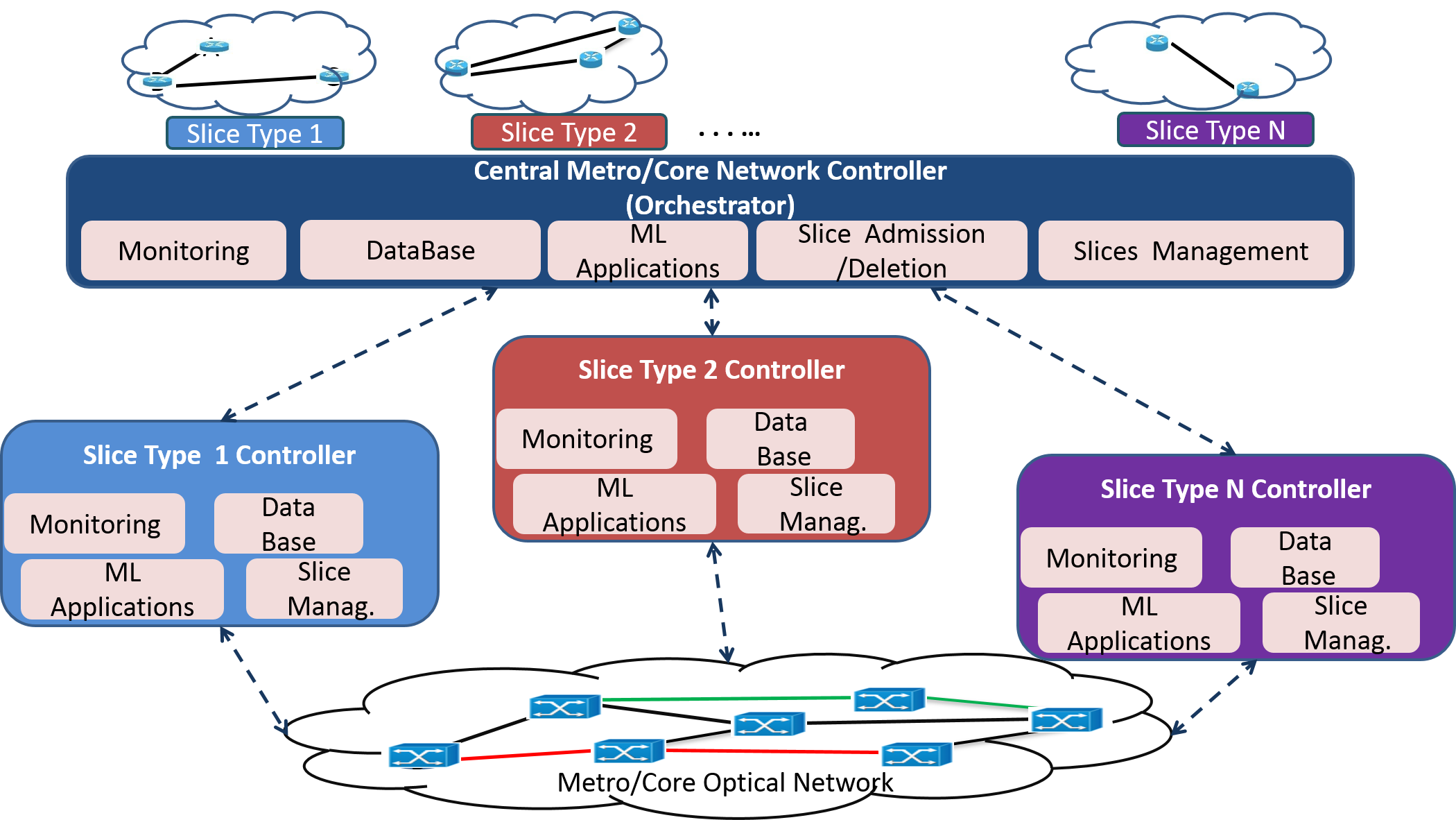}
\caption{Multi-slice Control and Management Framework}
\label{f2}
\end{center}
\end{figure}

For finding the QoT estimation model/s this work explores both centralized- and distributed-computed models, of the general multi-slicing control and management framework of Fig.~\ref{f2}. This framework consists of a central controller (orchestrator) and a number of distributed controllers (e.g., fog-computing based controllers), each with their own monitoring, storage, processing, and management capabilities.
    
The connection provisioning phase is performed centrally, according to global network information. Hence, the available network resources can be more efficiently managed. An ML-based QoT application, running on top of a database, extracts from its allocated database the appropriate type of data (i.e., BER data obtained via optical performance monitoring (OPM)) and infers from these data the QoT estimation model/s. According to Fig.~\ref{f2}, the model inference procedure can be performed either centrally (by the central controller) or in a distributed fashion (by the local controllers). In the first case, the necessary information is stored and processed centrally, whereas in the second case, each local controller is responsible for storing and processing information that is relevant only to a particular slice type (e.g., a slice that accommodates connections with the same BER requirement). By doing so, each slice type can be managed/controlled by its own QoT model that is inferred locally and independently from the other QoT models. The trainable parameters of the distributed QoT models are sent to the central controller for subsequently performing the QoT-aware connection/slice provisioning phase. 

In this work, the ML-based QoT estimation problem is formulated and compared according to the aforementioned centralized and distributed frameworks. Note that in the centralized case, the distributed controllers can be omitted from Fig.~\ref{f2}. However, for simplicity, we have chosen to illustrate both frameworks in a single figure.  

\subsection{Centralized QoT Formulation}\label{cent}
The centralized QoT estimation problem is formulated as a multiclass classification problem for an EON. Given a dataset $D=({\bf x},{\bf y})=\{{\bf x}^i, {\bf y}^i\}_{i=1}^N$ where, ${\bf x}^i$ is a vector capable of describing lightpath $i$, and ${\bf y}^i$ is the vector describing the BER ground truth of connection $i$, we are interested to infer from $D$, a QoT model $f({\bf x}^i)={\bf y}^i$ that accurately estimates the QoT class of lightpath $i$.  

The vector ${\bf x}^i=\{x_j^i\}_{j=1}^c$ contains information regrading $c$ features of lightpath $i$. Specifically,\\
$\bullet$ $x^i_{1}:$ is the entire length (in km) of lightpath $i$ \\
$\bullet$ $x^i_{2}:$ is the maximum link length (in km) of lightpath $i$ \\
$\bullet$ $x^i_{3}:$ is the central frequency allocated to lightpath $i$ \\ 
$\bullet$ $x^i_{4}:$ is the number of slots allocated to lightpath $i$ \\ 
$\bullet$ $x^i_{5}=1,2,3,4:$ if BPSK, QPSK, 8-QAM or 16-QAM modulation format is used for lightpath $i$, respectively \\ 
$\bullet$ $x^i_{6}:$ is the number of EDFAs along lightpath $i$ \\ 
$\bullet$ $x^i_{7}:$ is the number of links along lightpath $i$. 

The vector ${\bf  y}^i=\{y^i_j\}_{j=1}^{K+1}$ declares the QoT class of lightpath $i$; that is, if $y^i_j=1$, then  lightpath $i$ belongs to QoT class $j$. Note that $\sum_{j} {\bf y}^i=1$ and each element in ${\bf y}^i$ can take either the value one or zero, hence always indicating a single QoT class. In this work, the QoT classes are defined according to the BER requirements of all the possible slice types. Given the set of all possible BER requirements ${\bf B}=\{B_k\}_{k=1}^K$, where $K$ is the number of all the possible slice types, and $B_k < B_{k+1}$, then the following $K+1$ QoT classes can be defined according to the ground truth BER$^i$ of each lightpath $i$: \\
$\bullet$ Class $1$: If BER$^i < B_1$, then $y^i_1=1$ \\
$\bullet$ Class $v$: If $B_{v-1} \leq$ BER$^i < B_v$, then  $y^i_v=1$ $\forall 1<v\leq K$\\
$\bullet$ Class $K+1$: If BER$^i > B_K$, then $y^i_{K+1}=1$. 

In this work, the QoT model $f({\bf x}^i)={\bf y}^i$ is inferred according to a neural network (NN). After the inference procedure, the QoT model takes as input a vector ${\bf x}^i$, that has not been used during the training procedure, and returns a vector ${\bf y}^i$. If the BER requirement of lightpath $i$ is $B_j$ and the model returns $y^i_v=1$, then  lightpath $i$ is feasible (i.e., its BER$^i$ is lower than the $B_j$ requirement) if $v<j$, otherwise lightpath $i$ is infeasible.
\subsection{Distributed QoT Formulation}
The distributed QoT estimation problem is formulated according to a set of binary classifiers for an EON. Given a set of datasets ${\bf D}=\{D_k\}_{k=1}^K$, where $D_k$ is the dataset stored in the local controller of slice type $k$, we are interested to infer from ${\bf D}$ a set of QoT models ${\bf f}=\{f_k\}_{k=1}^K$, where $f_k$ is the QoT model of slice type $k$.   
 
Specifically, $D_k=({\bf x}_k,{y_k})=\{{\bf x}^{i}_k, {y}^{i}_k\}_{i=1}^{N'}$ where, ${\bf x}^{i}_k$ is the vector describing lightpath $i$ that is intended for slice type $k$, ${y}^{i}_k$ describes the BER ground truth of lightpath $i$, and $N'$ is the number of patterns in dataset $D_k$. The ${\bf x}^{i}_k$ vector is as previously defined in Section~\ref{cent}, with the difference that it refers only to the lightpaths that are established/indented for slice type $k$. The ${y}^{i}_k$ element can either be $0$ or $1$. Specifically, if ${y}^{i}_k=0$ then the lightpath $i$ of slice type $k$ is above its BER requirement $B_k$ (Class 1 - infeasible Class), and if ${y}^{i}_k=1$ (Class 2 - feasible Class) then the lightpath $i$ of slice type $k$ is below its BER requirement $B_k$. 

The set of QoT models, ${\bf f}$, is inferred in a distributed fashion, according to a set of neural networks (NNs) that are trained locally and independently. After the inference procedure, ${\bf f}$ is sent to the central controller and used during the QoT-aware connection provisioning phase. Each QoT model, $f_k$, takes as input a vector ${\bf x}^{i}_k$, that has not been used during the training procedure, and returns a ${y}^{i}_k$ value. If ${y}^{i}_k=1$, lightpath $i$ is feasible, otherwise it is declared as infeasible. 
\section{Dataset Generation}
\begin{figure}[h]
\begin{center}
\includegraphics[scale=0.23]{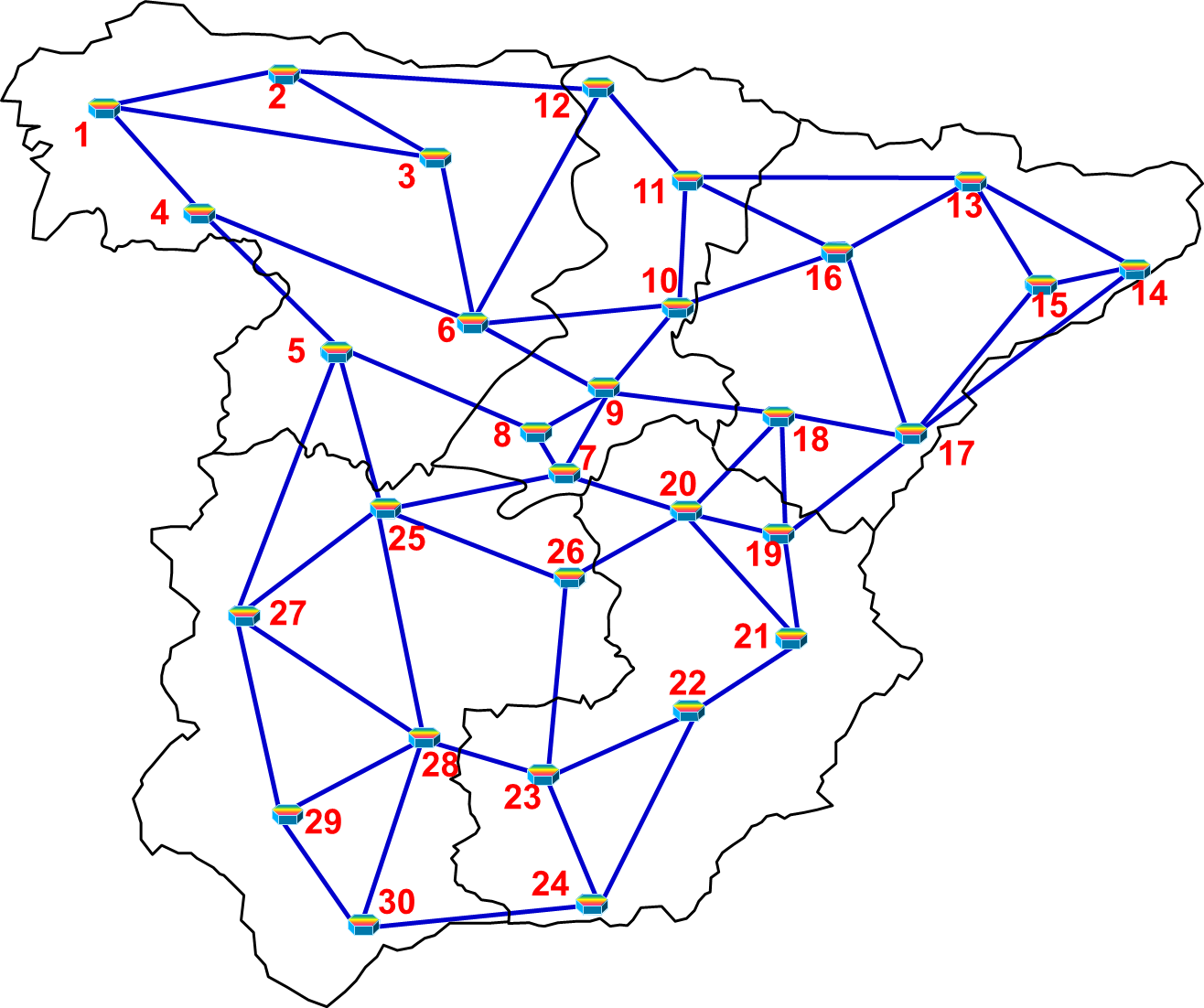}
\caption{Telefonica network topology}
\label{net}
\end{center}
\end{figure}

We assume an EON that is implemented using bandwidth variable transponders operating at multiple modulation formats: BPSK, QPSK, $8$-QAM, and $16$-QAM. Moreover, a flexible grid is implemented with channel spacing of $25$ GHz, with a baud rate equal to $16$ Gbaud, which results in an overall number of $160$ frequency slots for each link in the network. The national Telefonica network topology (Fig.~\ref{net}) was used, consisitng of $30$ nodes and $56$ undirected links.

In total, $N=20000$ connection requests were generated in a dynamic optical network according to a Poisson process with exponentially distributed holding times for a network load of $400$ Erlangs. The set of connection requests, ${\bf C}=\{C^i\}_{i=1}^{N}$, defining the source node $s$, the destination node $d$, and the BER requirement $B_k$ of each connection $i$ were generated as follows: The $s$-$d$ pairs were generated by randomly sampling from the set of network nodes; the $B_k$ requirement was generated by randomly sampling from a predefined set of possible BER requirements ${\bf B}=\{B_k\}_{k=1}^K$; a bit-rate was randomly generated for each connection request $C^i$ by uniformly sampling between the bit-rate interval that varies from $10$ to $200$ Gbps.

For each connection request $C^i$, a conventional routing and spectrum allocation (RSA) algorithm was used. Specifically the routing problem was solved according to Dijkstra's shortest path algorithm~\cite{Dijkstra+09} and the spectrum allocation problem was solved according to the first-fit scheme, with the aim of finding a lightpath that meets the spectrum continuity, contiguity, and no-frequency overlap constraints. The ground truth BER$^i$ for each computed lightpath was estimated according to the Q-tool described in~\cite{31}. Note that in actual implementations, probing lightpaths or alien wavelengths~\cite{17,36} can be used for enriching the dataset information (especially for the infeasible connections). Transfer learning~\cite{37} or active learning~\cite{38} techniques can be also used for reducing the number of probes required for finding accurate QoT models. 
 
\subsection{Centralized Dataset}
The multiclass dataset $D=\{{\bf x}^i, {\bf y}^i\}_{i=1}^N$ was created by extracting from each computed lightpath $C^i$ its ${\bf x}^i$ and ${\bf y}^i$ vectors. Vector ${\bf x}^i$ consists of the lightpath's features described in Section~\ref{cent}. Vector ${\bf y}^i$ was encoded according to the ground truth BER$^i$ (generated by the Q-tool~\cite{31}) and according to the set of BER requirements $\bf B$ defining the number of possible classes.  
\subsection{Distributed Datasets}
The datasets ${\bf D}=\{D_k\}_{k=1}^K$ were created by partitioning the mutliclass dataset $D$ into $K$ binary datasets as follows: Each dataset $D_k=\{{\bf x}^{i}_k, {y}^{i}_k\}_{i=1}^{N'}$ was created by extracting from $D$ only the patterns (connections) $i$ with a BER requirement that is equal to $B_k$. Feature vector ${\bf x}^{i}_k$ can then be found readily available in $D$, while ${y}^{i}_k$ is encoded to a binary value as follows: If the ground truth BER$^i$ of lightpath $i$ is below $B_k$ then ${y}^{i}_k=1$, otherwise ${y}^{i}_k=0$. 
\section{QoT Model Training}
For training the QoT models we applied a NN with $1$ hidden layer and $6$ hidden units. The rectified linear units (ReLU) function was used for the hidden layer that is followed by a softmax layer as the output layer. The crossentropy loss function was used that was optimized according to the ADAM~\cite{adam} algorithm. The learning rate was set to $0.01$, the number of training epochs was set to $300$, the batch size was set to $50$, and the size of the validation dataset was set to $20\%$ of the total number of patterns in the dataset (e.g., $20\%$ of the total patterns in $D$). We performed 3-fold cross validation and the model's accuracy was averaged over all the validation folds. The same procedure was followed for both the centralized and distributed datasets with the difference that in the centralized multiclass classification case the softmax layer has $K+1$ outputs (equal to the total number of classes in the dataset $D$), while in the distributed binary classification case the softmax layer has just $2$ outputs. Note that alternative ML techniques can also be applied. However, in this work we have opted for NNs as they have shown to present better generalization and higher accuracy than other ML techniques used for QoT estimation purposes~\cite{24}. Furthermore, the use of NNs has been also experimentally demonstrated for elastic optical networks~\cite{36}.

\section{QoT Model Evaluation}
After the training procedure, both models were evaluated and compared according to their accuracy and training time. The QoT models were trained and tested on a PC with a CPU $@$2.60 GHz and 8 GB RAM.   

\subsection{Centralized QoT model}\label{cent2}
For the centralized framework we have examined three different cases that vary according to the number of possible slice types. In the first case, we have assumed that the number of possible slice types is $K=3$ and the set of BER requirements is ${\bf B}=\{10^{-8}, 10^{-6}, 10^{-4}\}$. In the second case, we have assumed that $K=5$ with ${\bf B}=\{10^{-8}, 10^{-7}, 10^{-6}, 10^{-5},10^{-4}\}$, while in the third case we have assumed that $K=6$ with ${\bf B}=\{10^{-8}, 10^{-7}, 5\times 10^{-7}, 10^{-6}, 10^{-5},10^{-4}\}$. Note that these values were chosen so that the Q-tool~\cite{31} can be utilized to estimate the BER ground truth of each connection request.

According to the above, three QoT models were inferred that vary according to the number of classes. As described in Section~\ref{cent}, $K$ BER requirements (slice types) lead to $K+1$ classes. Hence, three classifiers with $4$, $6$, and $7$ classes were trained and tested. Table~\ref{tab1} summarizes the results regarding the models' accuracy, accuracy per class, and training time.

Results indicate that all three models achieve an overall high accuracy that drops as the number of classes increases. The models' accuracy per class is also high in most classes, however, as the number of classes increases it drops significantly, below an acceptable accuracy (i.e., drops to an accuracy of $47\%$ for the $7$ classes); a reasonable outcome if we consider that the model has to distinguish between a larger number of classes, negatively affecting the success rate of each class. Specifically, as the number of classes increases, the BER requirements of the different slice types become closer and thus the problem of distinguishing between the various classes becomes harder. As the training time mainly depends on the number of training patterns, which is the same for all three models ($|D|=20000$), it does not significantly change with the number of classes. 
    
\begin{table}[h!]
\centering
\caption{Centralized QoT Model}
\begin{tabular}{l|c|c|c}
          &\shortstack{4 Classes \\ ($K=3$)} & \shortstack{6 Classes \\ ($K=5$)} & \shortstack{7 Classes \\ ($K=6$)}  \\ \hline
Model Acc. ($\%$) & 96 & 93 & 91\\ \hline
Class 1 Acc. ($\%$) & 99 & 98 &  98\\ \hline
Class 2 Acc. ($\%$) & 96 & 95 & 94\\ \hline
Class 3 Acc. ($\%$) & 97 & 93 & 95\\ \hline
Class 4 Acc. ($\%$) & 90 & 94 & 47    \\ \hline
Class 5 Acc. ($\%$) & - & 91 & 91\\ \hline
Class 6 Acc. ($\%$) &  - & 88 & 94\\ \hline
Class 7 Acc. ($\%$) &  - & - & 97\\ \hline
Train. Time (sec) & 307 & 368 & 348 \\
\end{tabular}
\label{tab1}
\end{table}    

Overall, the results indicate that adoption of a centralized multiclass classification approach can be feasible only up to a specific number of diverse slice types. As this number may depend on several parameters (e.g., actual BER requirements, network technology utilized, etc.), its investigation is out of the scope of this work (field trials could be used). Beyond this number of diverse slice types, however, and as discussed next, the distributed QoT estimation framework can be adopted.    
\subsection{Distributed QoT models}
The distributed QoT framework was examined for the case where $K=5$ (5 slice types) with ${\bf B}=\{10^{-8}, 10^{-7}, 10^{-6}, 10^{-5},10^{-4}\}$ and for the case where $K=6$ with ${\bf B}=\{10^{-8}, 10^{-7}, 5\times 10^{-7}, 10^{-6}, 10^{-5},10^{-4}\}$. Note that these are the cases where the multiclass classifier has shown a poorer performance in Section~\ref{cent2}. In the first case, $5$ binary classifiers (slice types) were trained and in the second case $6$ binary classifiers were trained. Tables~\ref{tab2} and~\ref{tab3} summarize the results for the $5$ and $6$ binary classifiers,  respectively, regarding their overall accuracy, accuracy per class, and training time.

The results indicate that all QoT models achieve an overall high accuracy that is above $97\%$. Importantly, all the models achieve an overall high accuracy within each one of the two classes; the infeasible BER (Class $1$), and the feasible BER (Class $2$). The training time for all the models varies between $58$ and $178$ seconds. This is due to the fact that the number of patterns in each binary dataset may not be the same, affecting the time required for training each classifier. As previously described, each binary classifier $k$ is trained only according to the patterns/lightpaths indented for slice type $k$ (i.e., according to the lightpaths with a BER requirement $B_k$). In this work, for the $K=5$ case (Table~\ref{tab2}), $|D_1|=3349$, $|D_2|=4975$, $|D_3|=2057$, $|D_4|=4113$, and $|D_5|=5143$, where $|D_k|$ indicates the number of patterns in the dataset $D_k$. For the $K=6$ case (Table~\ref{tab3}), $|D_1|=2500$, $|D_2|=3340$, $|D_3|=4975$, $|D_4|=2579$, $|D_5|=7033$ and $|D_6|=1289$. Note that according to Tables~\ref{tab2} and~\ref{tab3} the training time increases as the number of patterns increases.

\begin{table}[h!]
\centering
\caption{Distributed QoT Models for $K=5$}
\begin{tabular}{c|c|c|c|c}
\shortstack{Slice Type \\ ($k$)} & \shortstack{Model \\ Acc. ($\%$)} & \shortstack{Class 1 \\ Acc. ($\%$)} & \shortstack{Class 2 \\ Acc. ($\%$)} & \shortstack{Train. Time \\ (sec)}  \\ \hline
1 & 99 & 100 & 99 & 61 \\ \hline
2 & 98 & 97 & 99 & 84  \\ \hline
3 & 99 & 98 & 99 & 98 \\ \hline
4 & 97 & 95 & 98 & 80 \\ \hline
5 & 98 & 98 & 95 & 100 \\ \hline
\end{tabular}
\label{tab2}
\end{table}

\begin{table}[h!]
\centering
\caption{Distributed QoT Models for $K=6$}
\begin{tabular}{c|c|c|c|c}
\shortstack{Slice Type \\ ($k$)} & \shortstack{Model \\ Acc. ($\%$)} & \shortstack{Class 1  \\ Acc. ($\%$)} & \shortstack{Class 2 \\ Acc. ($\%$)} & \shortstack{Train. Time \\ (sec)}  \\ \hline
1 & 99 & 95 & 99 & 64 \\ \hline
2 & 100 & 100 & 100 & 87  \\ \hline
3 & 97 & 98 & 97 & 119 \\ \hline
4 & 98 & 97 & 98 & 77 \\ \hline
5 & 97 & 98 & 97 & 178 \\ \hline
6 & 97 & 98 & 94 & 58 \\ \hline
\end{tabular}
\label{tab3}
\end{table}

Overall, the distributed QoT estimation framework greatly outperforms the centralized QoT estimation framework in all metrics examined; that is the models' accuracy, accuracy per class, and training time. Specifically, the training time achieved by the distributed framework outperforms the training time achieved by the centralized framework by approximately $50\%$. This is due to to the fact that the number of patterns in the distributed controllers (that keep only local, per service type information) is significantly lower than the number of patterns in the centralized controller (that keeps global information). Evidently, the distributed framework, requires less processing and storage resources per slice type controller and offloads the central controller from functionalities that can be effectively performed in distributed controllers. Note that investigating the resource allocation problem (storage, processing, bandwidth) as well as the control overhead for each approach examined, is out of the scope of this work and is planned for future work.        

Importantly, the results show that while the accuracy of the centralized case is negatively affected by the number of service types, the accuracy of the distributed case is not. In particular, when the distributed approach is followed, the models' accuracy per class is consistently above $94\%$, while the accuracy per class for the centralized approach is above $90\%$ for the smallest set of diverse QoT requirements and falls to around $47\%$ for the largest set of diverse QoT requirements examined). In the distributed case, the QoT model is always a binary classifier, and hence, no matter the number of slice types, it will always have to distinguish between two classes. In the centralized case, however, the QoT model is a multiclass classifier that depends on the number of slice types. Hence, as the number of slice types increases, the more difficult it becomes for the classifier to distinguish between the classes, a fact that indicates that as the diversity of QoT requirements increases the centralized model can no longer be utilized, as its accuracy becomes unacceptably low.    

\section{Conclusions}
Centralized and distributed QoT estimation frameworks are examined for optical networks supporting slices with diverse BER requirements. A multiclass classifier is formulated for the centralized framework and a set of binary classifiers is formulated for the distributed framework. All QoT models were trained according to a NN. Results demonstrate that the accuracy per class of the centralized QoT model is negatively affected as the number of diverse slice types increases (drops to $47\%$). Distributed QoT models, however, being independent from the number of diverse slice types, attain an overall high accuracy in both classes of interest (above $94\%$). Furthermore, as each distributed QoT model is inferred only according to the lightpaths intended for the same slice type, the training time for each distributed QoT model is significantly decreased (by approximately $50\%$) when compared to the training time of the centralized QoT that is inferred according to global network information. For future work we are planning to examine ML-based centralized and distributed frameworks for the efficient control and management of optical network slices that consider, apart from the diverse QoT requirements, additional QoS requirements (i.e., diverse bit-rate requirements that vary over time).                     

\section*{Acknowledgment}
This work has been supported by the European Union’s Horizon 2020 research and innovation programme under grant agreement No 739551 (KIOS CoE) and from the Government of the Republic of Cyprus through the Directorate General for European Programmes, Coordination and Development.




\begin{thebibliography}{1}


\bibitem{1}
5GPPP Architecture Working Group, ``View on 5G Architecture,'' 2017.

\bibitem{2}
S. Kavanagh, ``What is Network Slicing?,'' 5G.co.uk, online: https://5g.co.uk/guides/what-is-network-slicing/, 2018.

\bibitem{3}
I. Scales, ``Nokia Claims Network Slicing for the Fixed Network,'' https://www.telecomtv.com/content/fixed-access/nokia-claims-network-slicing-for-the-fixed-network-32703/, 2018.

\bibitem{4}
A. Farrel, ``Service Function Chaining (SFC) and Network Slicing in Backhaul and Metro Networks in Support of 5G,'' \emph{Proc. IEEE ICTON}, 2018.

\bibitem{5}
SNS: Market Intelligence and Consultancy Solutions Res., ``SON (Self-Organizing Networks) in the 5G Era: 2019–2030 – Opportunities, Challenges, Strategies $\&$ Forecasts,'' 2018.

\bibitem{6}
A. Mayoral, et al., ``Multi-tenant 5G Network Slicing Architecture with Dynamic Deployment of Virtualized Tenant Management and Orchestration (MANO) Instances,'' \emph{Proc. ECOC}, 2016.

\bibitem{7}
M. R. Raza, et al., ``Demonstration of Resource Orchestration Using Big Data Analytics for Dynamic Slicing in 5G Networks,'' \emph{Proc. ECOC}, 2018.

\bibitem{8}
R. Alvizu, et al., ``Network Orchestration for Dynamic Network Slicing for Fixed and Mobile Vertical Services,'' \emph{Proc. IEEE/OSA OFC}, 2018.

\bibitem{12}
J. Mata, et al., ``Artificial Intelligence (AI) Methods in Optical Networks: A Comprehensive Survey,'' \emph{Opt. Switc. Net.}, 28:43--57, 2018.

\bibitem{13}
F. Musumeci, et al, ``An Overview on Application of Machine Learning Techniques in Optical Networks,'' \emph{IEEE Com. Surveys $\&$ Tutorials}, 21(2):1383--1408, 2019.


\bibitem{17}
X. Chen, et al., ``Knowledge-Based Autonomous Service Provisioning in Multi-Domain Elastic Optical Networks,'' \emph{IEEE Comm. Mag.}, 56(8): 152--158, 2018.

\bibitem{18}
R. Alvizu, et al., ``Matheuristic with Machine-learning-based Prediction for Software-defined Mobile Metro-core Networks,''  \emph{IEEE/OSA J. Opt. Comm. Net.}, 9(9):D19--D30, Sept. 2017.

\bibitem{19}
F. Morales, et al., ``Virtual Network Topology Adaptability based on Data Analytics for Traffic Prediction,''  \emph{IEEE/OSA J. Opt. Comm. Net.}, 9(1):A35–-A45, 2017.

\bibitem{20}
T. Panayiotou, et al., ``On Learning Bandwidth Allocation Models for Time-Varying Traffic in Flexible Optical Networks,'' \emph{Proc. IEEE ONDM}, 2018.

\bibitem{21}
T. Panayiotou, et al., ``A Data-driven Bandwidth Allocation Framework with QoS Considerations for EONs,'' \emph{IEEE/OSA J. Light. Techn.}, 37(9):1853--1864, 2019.


\bibitem{26}
T. Panayiotou et al., ``Leveraging Statistical Machine Learning to Address Failure Localization in Optical Networks,'' \emph{IEEE/OSA J. Opt. Comm. Net.}, 10(3):162--173, 2018.

\bibitem{27}
B. Shariati, et al., ``Learning from the Optical Spectrum: Failure Detection and Identification'', \emph{IEEE/OSA J. Light. Techn.}, 37(2):433--440, 2019.

\bibitem{28}
S. Shahkarami, et al., ``Machine-Learning-Based Soft-Failure Detection and Identification in Optical Networks,'' \emph{Proc. IEEE/OSA OFC}, 2018.

\bibitem{28b}
B. Yan, et al., ``First Demonstration of Imbalanced Data Learning-Based Failure Prediction in Self-Optimizing Optical Networks with Large Scale Field Topology,'' \emph{Proc. ACP} 2018.

\bibitem{28c}
C. Natalino, et al., ``Field Demonstration of Machine-Learning-Aided Detection and Identification of Jamming Attacks in Optical Networks,'' \emph{Proc. ECOC}, 2018. 

\bibitem{22}
T. Panayiotou, et al., ``Performance Analysis of a Data-driven Quality-of-transmission Decision Approach on a Dynamic Multicast-capable Metro Optical Network,''  \emph{IEEE/OSA J. Opt. Comm. Net.}, 9(1):98--108, 2017.


\bibitem{23}
T. Panayiotou, et al., ``Machine Learning for QoT Estimation of Unseen Optical Network States,'' \emph{Proc. IEEE/OSA OFC}, 2019. 

\bibitem{24}
R. Morais and J. Pedro, ``Machine Learning Models for Estimating Quality of Transmission in DWDM Networks,'' \emph{IEEE/OSA J. Opt. Comm. Net.}, 10(10):D84--D99, 2018.


\bibitem{25}
J. Mata et al., ``Supervised Machine Learning Techniques for Quality of Transmission Assessment in Optical Networks,'' \emph{Proc. IEEE ICTON}, 2018.

\bibitem{25b}
C. Rottondi, et al., ``Machine-learning Method for Quality of Transmission Prediction of Unestablished Lightpaths,'' \emph{IEEE/OSA J. of Opt. Comm. Net.}, 10(2):A286--A297, 2018.

\bibitem{15b}
W. Fawaz, et al., ``Service Level Agreement and Provisioning in Optical Networks,'' \emph{IEEE Comm. Mag.}, 42(1):36--43, 2004.

\bibitem{15}
S. Yan, et al., ``Data-driven Network Analytics and Network Optimisation in SDN-based Programmable Optical Networks,'' \emph{Proc. IEEE ONDM}, 2018.





\bibitem{Dijkstra+09}
T. H. Cormen, C. E. Leiserson, R. L. Rivest, C. Stein, \emph{Introduction to Algorithms}, ``Section 24.3: Dijkstra's algorithm'', MIT Press, 2009.

\bibitem{31}
B. Shariati, et al., ``Physical-Layer-Aware Performance Evaluation of SDM Networks Based on SMF Bundles, MCFs, and FMFs,'' \emph{IEEE/OSA J. Opt. Comm. Net.}, 10(9):712--722, 2018.










\bibitem{36}
R. Proietti, et al., ``Experimental Demonstration of Cognitive Provisioning and Alien  Wavelength Monitoring in Multi-domain EON'', \emph{Proc. IEEE/OSA OFC}, 2018.

\bibitem{37}
W. Mo, et al., ``ANN-Based Transfer Learning for QoT Prediction in Real-Time Mixed Line-Rate Systems,'' \emph{Proc. IEEE/OSA OFC}, 2018.

\bibitem{38}
D. Azzimonti, et al., ``Using Active Learning to Decrease Probes for QoT Estimation in Optical Networks,'' \emph{Proc. IEEE/OSA OFC}, 2019.

\bibitem{adam}
D. Kingma and J. Ba, ``Adam: A Method for Stochastic Optimization,'' \emph{Proc. ICLR}, 2015.


\end{thebibliography}
%

\end{document}